\documentclass[aps,prl,twocolumn,superscriptaddress,showpacs]{revtex4}
\usepackage{amsmath,amssymb}

\usepackage{graphicx}

\begin{document}


\title{Transmission-phase measurement of the 0.7 anomaly in a quantum point contact}


\author{Toshiyuki Kobayashi}
\affiliation{NTT Basic Research Laboratories, NTT Corporation, Atsugi-shi, Kanagawa 243-0198, Japan}
\author{Shoei Tsuruta}
\affiliation{NTT Basic Research Laboratories, NTT Corporation, Atsugi-shi, Kanagawa 243-0198, Japan}
\affiliation{Tokyo University of Science, Shinjuku-ku, Tokyo 162-8601, Japan}
\author{Satoshi Sasaki}
\affiliation{NTT Basic Research Laboratories, NTT Corporation, Atsugi-shi, Kanagawa 243-0198, Japan}
\author{Hiroyuki Tamura}
\affiliation{NTT Basic Research Laboratories, NTT Corporation, Atsugi-shi, Kanagawa 243-0198, Japan}
\author{Tatsushi Akazaki}
\affiliation{NTT Basic Research Laboratories, NTT Corporation, Atsugi-shi, Kanagawa 243-0198, Japan}
\affiliation{Tokyo University of Science, Shinjuku-ku, Tokyo 162-8601, Japan}


\date{\today}

\begin{abstract}
We measure the transmission phase of a quantum point contact (QPC) at a low carrier density in which electron interaction is expected to play an important role and anomalous behaviors are observed. In the first conductance plateau, the transmission phase shifts monotonically as the carrier density is decreased by the gate voltage. When the conductance starts to decrease, in what is often called the 0.7 regime, the phase exhibits an anomalous increase compared with the noninteracting model. The observation implies an increase in the wave vector as the carrier density is decreased, suggesting a transition to a spin-incoherent Luttinger liquid.
\end{abstract}

 

\maketitle


A quantum point contact (QPC) is a tunable quasi-one-dimensional (1D) conductor fabricated in a semiconductor heterostructure \cite{QPC_vanWees1988,QPC_Wharam1988}. A clean QPC has been an important building block of quantum devices, as well as an ideal test bed for exploring the physics of 1D conductors. The conductance $G$ of a QPC is quantized in units of $2e^2/h$ and can be tuned by controlling the gate voltage. In a noninteracting picture, the conductance remains constant within the plateau and decreases monotonically when the conduction channel depletes. However, a shoulder-like structure around $0.7 \times 2e^2/h$ has been routinely observed prior to the depletion and the anomalous features are believed to be caused by a many-body spin-related phenomenon \cite{07_Thomas1996, 07_CEO2004, 07_GaN2005, 07_Hole2008, 07_InGaAs2008}. Although various experimental and theoretical studies have been undertaken in an attempt to explain this anomalous behavior, the origin remains unclear.

A compelling model explaining 0.7 features has been proposed by Matveev in the context of a spin-incoherent Luttinger liquid (SILL), in which the $2e^2/h$ plateau is suppressed to $e^2/h$ because the collective spin mode cannot propagate through the QPC because the exchange energy $J$ becomes smaller than $k_BT$ at a low electron density \cite{SILL_Matveev_PRL2004, SILL_Matveev_PRB2004, SILL_Matveev2009}. Here, $k_B$ is Boltzmann's constant and $T$ is the electronic temperature. Hew \textit{et al.} have reported the suppression of the $2e^2/h$ plateau to $e^2/h$ using a weakly confined QPC and shown the possibility of realizing a SILL regime in a low-electron-density QPC \cite{SILL_Hew2008}. However, concrete evidence for a SILL in the 0.7 regime remains to be found. 

In this Letter, we measure the transmission-phase of a QPC and show that the phase shift also exhibits anomalous behavior in the 0.7 regime. The phase showed a marked increase compared with the noninteracting model when the carrier density is decreased from the first conductance plateau. We explain this anomalous phase increase in terms of the transition from a Luttinger liquid (LL) to an SILL at a low carrier density. In the SILL, the 1D wave vector $k$ responsible for the transmission phase doubles from the noninteracting 1D Fermi wave vector $k_F$ to $2k_F$ \cite{SILL_Cheianov2004, SILL_Fiete2004, SILL_Cheianov2005, SILL_Fiete2005, SILL_Fiete2007, SILL_Fiete2010}. The observed phase anomaly provides additional evidence for the SILL model of the 0.7 anomaly. 

\begin{figure} [b]
\begin{center}
\includegraphics{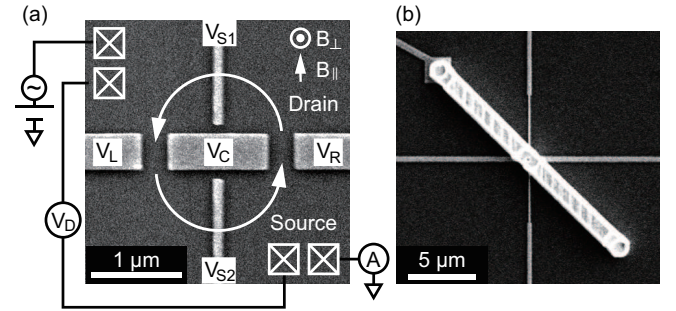}
\end{center}
\noindent
\caption{(a) SEM image of gate electrodes on the sample surface and the measurement setup. The arrow next to $B_{||}$ indicates the parallel component of the magnetic field when the device is tilted. (b) SEM image of the bridge electrode used to connect the island gate electrode and apply $V_C$ in (a).}
\label{fig:Fig1.pdf}
\end{figure}

The device used in this experiment was fabricated on a GaAs/AlGaAs heterostructure with a two-dimensional electron gas (2DEG) located 80 nm below the surface. The carrier density and mobility of the 2DEG estimated by Hall measurement at $T$ = 4 K were $n_{2D} = 2.2\times 10^{15}$ m$^{-2}$ and $\mu = 250$ m$^2$V$^{-1}$s$^{-1}$, respectively. The Fermi energy $E_{F,2D} = 7.8$ meV and the effective mass $m^* = 0.067m_0$ of GaAs yield a Fermi wave vector of 2DEG, $k_{F,2D}$, of $1.2\times 10^8$ m$^{-1}$. Several Ti/Au gate electrodes were patterned using electron-beam lithography and lift-off processes [Fig.\ref{fig:Fig1.pdf}(a)] to fabricate two QPCs aligned in parallel. To apply a gate voltage to a center island gate electrode, a bridge electrode \cite{airbridge1994} was fabricated 370 nm above the surface [Fig.\ref{fig:Fig1.pdf}(b)]. The lithographic width $W$ and length $L$ of the QPCs were 300 and 400 nm, respectively, and the spacing between two QPCs, $D$, was 1.4 $\mu$m.

The measurements were carried out in a dilution refrigerator at $T=200$ mK. We applied negative gate voltages to $V_L$, $V_C$, and $V_R$ to form two parallel QPCs. Throughout the experiment, we applied $-0.48$ V to $V_C$ and +0.5 V to $V_{S1}$ and $V_{S2}$. To measure the differential conductance $G \equiv dI_D/dV_D$ of parallel QPCs, we applied an ac (73 Hz) drain voltage $V_D$ of 10 $\mu$V and measured the drain current $I_D$ using the standard lock-in technique. Both QPCs exhibited clear conductance plateaux at integer multiples of $2e^2/h$ and a 0.7 structure.

\begin{figure} [tb]
\begin{center}
\includegraphics{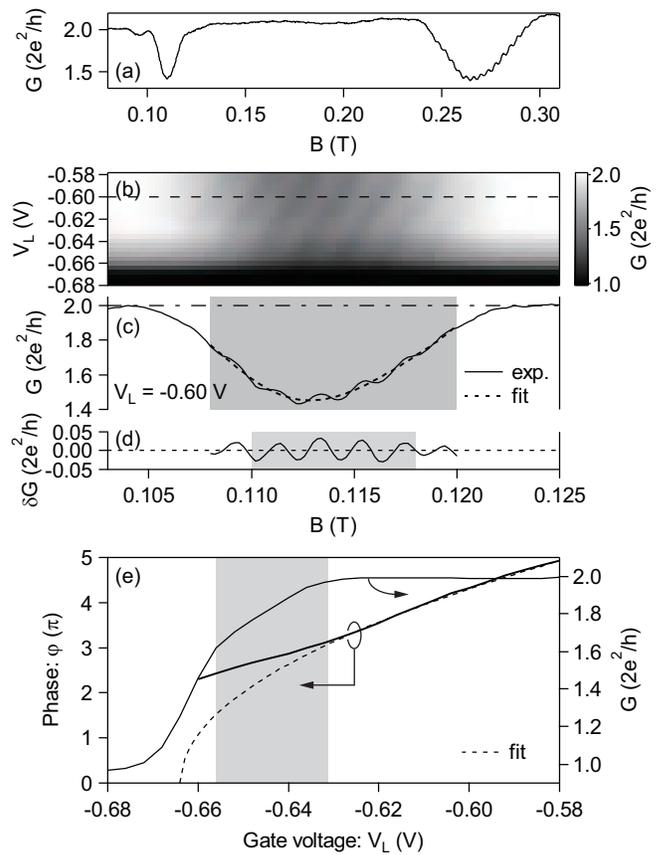}
\end{center}
\noindent
\caption{(a) Magnetic focusing in parallel QPC device. Both QPCs are tuned in the middle of the first conductance plateau. (b) Gray scale plot of $G$ under the first focusing condition. Sweeping both $B$ and $V_L$ result in stripe-pattern oscillations. (c) $G$ at $V_L=-0.60$ V and the polynomial fit. The shaded region was used for the fit. (d) Oscillatory component $\delta G$ extracted by subtracting the polynomial fit. The shaded region was used for the FFT to extract the phase.  (e) Transmission phase of the left QPC extracted by FFT at each $V_L$. The broken line is a square-root fit in the first plateau. The phase is shifted vertically to put it on the square-root fit. $G$ at an unfocused condition ($B=0.085$ T) is also plotted. The shaded region indicates the 0.7 regime.}
\label{fig:Fig2.pdf}
\end{figure}

To measure the transmission phase, we formed an interference loop by connecting two QPCs using cyclotron trajectories \cite{pQPC1989, pQPC_airbridge2000}. At magnetic fields $B$, where the integer multiples of the cyclotron diameter, $d_c=2\hbar k_{F,2D}/(eB)$, equal the spacing between two QPCs, electrons ejected from a QPC are focused on the other QPC \cite{vanHouten1989} and the total conductance decreases. Here, $B$ is aligned perpendicular to the 2DEG ($B = B_{\perp}$). By sweeping $B$ while both QPCs were tuned to the middle of the first plateau, two focusing signals were observed as dips in $G$ at $B$ = 0.114 T and 0.272 T [Fig.\ref{fig:Fig2.pdf}(a)]. The dip at $B$ = 0.272 T corresponds to the trajectory bouncing in the vicinity of the center gate electrode before entering the other QPC. Aharonov-Bohm (AB) oscillations were observed superposed on the large dips, reflecting the modulation of the phase acquired in the cyclotron trajectories through two QPCs. 

The transmission phase shift $\varphi$ in the QPC on the left was extracted from oscillations on $G$ by sweeping both $B$ and $V_L$, while keeping the QPC on the right at a fixed gate voltage in the middle of the first conductance plateau [Fig.\ref{fig:Fig2.pdf}(b)]. The oscillation exhibited a stripe pattern reflecting the modulation of the transmission phase acquired in the left QPC. The large background conductance dip caused by magnetic focusing was eliminated by fitting with a polynomial to extract the oscillatory component, $\delta G$ [Fig.\ref{fig:Fig2.pdf}(c) and (d)]. 

Fig.\ref{fig:Fig2.pdf}(e) shows the transmission phase shift induced by $V_L$ obtained by the fast Fourier transform (FFT) of the oscillatory component in Fig.\ref{fig:Fig2.pdf}(b) around $B$ = 0.112 T. Notice that the phase displayed in Fig.\ref{fig:Fig2.pdf}(e) does not represent an accurate transmission phase acquired in the QPC, since the FFT only provides information on the relative phase shift caused by $V_L$. Here, the small changes in the separation between the two QPCs induced by sweeping $V_L$ has negligible influence on the transmission phase shift measurement near the pinch-off of the QPC, because the phase acquired by the cyclotron loop is fixed by $B$ and $k_{F,2D}$. An analysis of the AB oscillation period revealed that the variation of the QPC length is less than 3\% for the gate voltage explored in our measurement.

In the first conductance plateau, $\varphi$ gradually decreased by sweeping $V_L$ toward the pinch-off. In the noninteracting model, $k_F$ is expected to show a square-root dependence on $V_L$, because negatively increasing $V_L$ causes a proportional reduction in the 1D Fermi energy, $E_F$, and $k_F \propto \sqrt{E_F}$. Although $k_F$ in the QPC and the interface between QPC and 2DEG are not uniform due to the saddle-like potential landscape, a square-root function, $\varphi \propto \sqrt{V_L-V_{L0}}$, fits well in the plateau [Fig.\ref{fig:Fig2.pdf}(e)]. Here, $V_{L0}$ is the $V_L$ value at the pinch-off. However, as $G$ is decreased from the plateau, at the transition to the 0.7 regime, the phase deflects from the square-root trend and exhibits anomalous an slow-down, which is not expected in the noninteracting model.

\begin{figure} [tb]
\begin{center}
\includegraphics{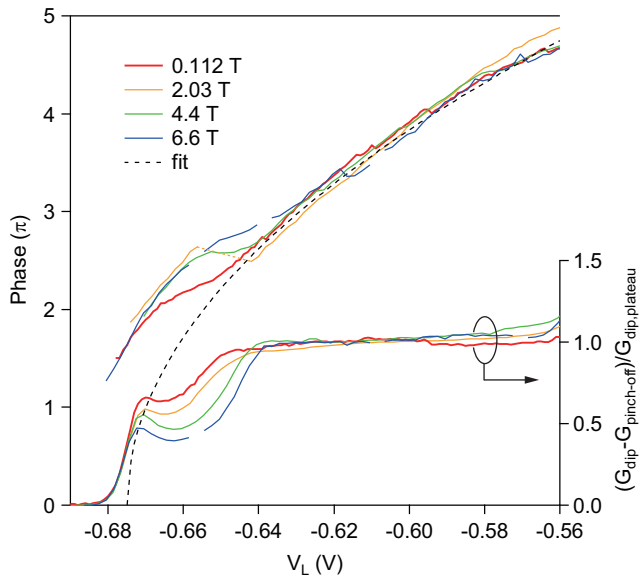}
\end{center}
\noindent
\caption{Transmission phases and conductance steps of the left QPC for the second cool-down. Each curve is horizontally shifted to align it at a pinch-off of 0.112 T. Phases are vertically shifted to put them on the curve of the square-root fit. Since $G$ at the magnetic focusing dip is decreased depending on the magnetic focusing efficiency and also modulated by AB oscillations, minimum conductance $G_{\textrm{dip}}$ is extracted by the polynomial fit of the focusing dip. Then, the component of the right QPC is eliminated by subtracting $G$ at the pinch-off, $G_{\textrm{pinch-off}}$. To compare conductance steps at different $B$ values where the focusing efficiencies are slightly different, $(G_{\textrm{dip}}-G_{\textrm{pinch-off}})$ was normalized by $G$ at the first plateau, $G_{\textrm{dip,plateau}}$. Notice that the normalized conductance indicates the $V_L$ at which the transmission starts to decrease; however, it does not represent the precise transmission in the QPC. Some data points missing from the curves at $B$ = 6.6 T are due to the noise during the measurement. The phase at $B$ = 2.03 T could not be extracted in the vicinity of $V_L$ = -0.65 V because of the weak oscillation amplitude. In this region, a broken line is shown to indicate the anomalous phase shift.}
\label{fig:Fig3.pdf}
\end{figure}

To test the reproducibility of the anomalous phase shift, we warmed the device to room temperature and then cooled it down again. By doing so, the randomly frozen charged impurities form a slightly different potential landscape and we can measure a QPC with different characteristics using the same device. Although the second cool-down showed different conductance characteristics in the 0.7 regime, we observed a similar phase shift anomaly (Fig.\ref{fig:Fig3.pdf}). This time, the improved oscillation amplitude allowed us to extract transmission phases at conductances below the 0.7 regime, where the conductance rapidly decreases. In this pinch-off regime, the phase shift recovered a strong dependence on $V_L$, indicating that the deviated phase shift in the 0.7 regime is anomalous behavior. 

To further investigate the phase shift anomaly, we measured the magnetic field dependence (Fig.\ref{fig:Fig3.pdf}). We tilted the device \textit{in situ} to apply a large $B$ while keeping $B_{\perp}$ small and thus explore oscillations using the same cyclotron trajectories. We measured the phase shift at four different tilt angles represented by the magnetic fields 0.112, 2.03, 4.4, and 6.6 T. By increasing $B$, the spin degeneracy of the 1D subband is gradually lifted and the conductance in the 0.7 regime is decreased due to increased reflection at the higher spin-subband. At a high $B$, the conductance starts to decrease at a higher carrier density (i.e. a less negative $V_L$), which results in a clear half-plateau. The normalized conductance steps for four $B$'s at which we measured the phase shift are also shown in Fig.3. The phase shift was extracted in a similar manner as described above for perpendicular $B$. Here, we extracted the transmission phase from approximately two oscillation periods of $\delta G$ to minimize the $B$ variation during the measurement, because the shift of the Zeeman energy during the measurement may influence the analysis of the phase shift. For instance, at 6.6 T, the FFT was performed for oscillations obtained between 6.47 and 6.7 T. As shown in Fig.\ref{fig:Fig3.pdf}, we still observe the anomalous phase deviation even at 6.6 T. It is also important to note that the onset of the phase anomaly shifts with increasing $B$.

To explain the phase shift anomaly, we considered several different models for the 0.7 anomaly including spontaneous spin polarization \cite{07_Thomas1996, sp_Rohkinson2006, sp_Graham2007, sp_Sfigakis2008, sp_Chen2008, sp_Chen2009, sp_Chen2010}, the Kondo effect \cite{Kondo_Cronenwett2002, Kondo_Rejec2006}, and SILL. In the spontaneous spin polarization model, a higher spin subband is considered to deplete slowly with the gate voltage near its pinch-off, which may result in a slow-down in the phase shift. However, the higher spin subband carries less current as it approaches the pinch-off, and its contribution to the conductance oscillations is smaller than the lower spin subband, in which the phase shift is not expected to exhibit anomalous behavior. Therefore, it is unlikely that the observed phase shift is dominated by the pinning of the higher spin subband. The Kondo effect caused by the formation of a quasi bound state within the QPC may also cause a phase anomaly because the transmission phase through the Kondo state is locked to $\pi /2$ at the unitarity limit \cite{kondophase_2000, kondotheory_2000, Kondophase_unitary_2002, kondophase_single_2008}. However, the phase shift anomaly is observed even at a high $B$ at which the Kondo effect should be substantially suppressed.

In the SILL model, the electronic state in the QPC gradually changes from a Luttinger liquid (LL) to an SILL as the 1D carrier density, $n$, is decreased by $V_L$. While the conductance of a short LL attached to noninteracting leads does not depend on the interaction strength, that of the SILL is expected to show a marked dependence on the electron interaction within the wire. As $n$ is decreased and the Coulomb interaction between adjacent electrons becomes larger than their kinetic energy, the probability of exchanging adjacent electrons decreases and $J$ becomes significantly lower than $k_BT$. In this SILL regime of $J \ll k_BT \ll E_F$, the conductance is suppressed from $2e^2/h$ because spinons with energies larger than $\pi J/2$ cannot propagate through the QPC \cite{SILL_Matveev2009, SILL_MonteCarlo2007}. Therefore, the 0.7 conductance anomaly can be seen as the transition from LL to SILL. Upon transition, the wave vector responsible for the electron transport doubles from $k_F$ of LL to 2$k_F$ of SILL, as has been shown in calculations of the spectral function in previous studies \cite{SILL_Cheianov2004, SILL_Fiete2004, SILL_Cheianov2005, SILL_Fiete2005, SILL_Fiete2007, SILL_Fiete2010}. If the entire QPC undergoes the transition to an SILL, the transmission phase, $\varphi_t = \int k \, dx$, doubles from the non-interacting model. In a real device, the potential is saddle shaped and $n$ gradually increases toward the 2DEG. Therefore, the length of the SILL region gradually extends as $n$ in the QPC is decreased by making $V_L$ more negative and the phase gradually deviates from the square-root trend. We also observe that the deviated phases again converge to the same trend even at different $B$ values. This is probably because the development of the SILL region slows markedly at a certain length due to the finite length of the QPC formed by a pair of rectangular gate electrodes. Once the SILL region has extended throughout the QPC, the transmission phase becomes the sum of the phases acquired in the SILL region and the interface region between QPC and 2DEG, resulting in the recovery of a strong phase shift toward the pinch-off. Taking all of these observations into account, we attribute the anomalous deviation in transmission phase to the development of the SILL region.

As $B$ is increased, the onset of the phase shift deviation, as well as the suppression in the conductance, moves toward a less negative $V_L$. This is explained by the strong dependence of SILL on $n$. When the conductance drops from the first plateau, the higher spin subband starts to deplete and $n$ rapidly decreases, triggering the formation of the SILL. When the higher spin subband starts to deplete at a low $B$, $n$ in the lower spin subband is much lower than that at a high $B$; therefore, SILL is expected to extend throughout the QPC more rapidly as the higher spin subband starts to deplete. In Fig. 3, the phase at 2.03 T \textit{increases} with a negatively increasing $V_L$, in contrast to the constant and weakly decreasing phase shifts at 4.4 and 6.6 T, respectively. If the additional phase acquisition by development of SILL ($k_F$ to $2k_F$) is larger than the decrease in $k_F$, the phase increases as in the case of 2.03 T. If the phase acquisition equals the decrease in $k_F$, the phase stays constant as in the case of 4.4 T. The phases at different $B$ values converge at nearly the same $V_L$ ($\approx -0.66$ V in Fig. 3) except for that of 0.112 T. This is probably because the Zeeman energy at 0.112 T is smaller than $k_BT$ and the spin subbands are degenerate, resulting in a slower development of SILL.

In summary, we have demonstrated the transmission phase measurement of a QPC. The transmission phase showed a marked deviation compared with the noninteracting model as the conductances started to decrease from the first plateau, including both the 0.7 regime at a low magnetic field and a transition to a half plateau at a high magnetic field. The anomalous phase shift was explained by the formation of a spin-incoherent Luttinger liquid in which the wave vector is twice that of a Luttinger liquid.

We thank T. Kubo, Y. Tokura, K. Muraki, T. Fujisawa, S. Miyashita, and T. Maruyama for fruitful discussions and technical support.

\end{document}